\documentclass[preprintnumbers,prd,nofootinbib,floats,amssymb,floatfix]{revtex4}

\usepackage{amsfonts}
\usepackage{amsmath}
\usepackage{graphicx}
\usepackage{color}
\usepackage{amssymb}
\usepackage{hyperref}
\hypersetup{colorlinks,linkcolor={magenta},citecolor={magenta},urlcolor={magenta}}
\usepackage[normalem]{ulem}

\begin{document}

\title{Conserved laws and dynamical structure of axions coupled to photons}

\author{Omar Rodr{\'i}guez-Tzompantzi}
\email{omar.tzompantzi@unison.mx}
%\email{omar.tz2701@gmail.com}
\affiliation{ Departamento de Investigaci\'on en F\'isica, Universidad de Sonora, Apartado postal 14740, C.P. 83000, Hermosillo, Sonora, M\'exico}
\affiliation{ Departamento de Ciencias F\'isicas, Universidad Andres Bello, Sazi\'e 2212, Piso 7, Santiago, Chile.}

%%%%%%%%%%%%%%%%%%%%% Abstract
\begin{abstract}
In this work, we carry out a study of the conserved quantities and dynamical structure of the four-dimensional modified axion electrodynamics theory described by the axion-photon coupling. In the first part of the analysis, we employ the covariant phase space method to construct the conserved currents and to derive the Noether charges associated with the gauge symmetry of the theory. We  further derive the improved energy-momentum tensor using the Belinfante-Rosenfeld procedure, which leads us to the expressions for the energy, momentum, and energy flux densities. Thereafter, with the help of Faddeev-Jackiw's Hamiltonian reduction formalism, we obtain the relevant fundamental brackets structure for the dynamic variables and the functional measure for determining the quantum transition amplitude. We also confirm that modified axion electrodynamics has three physical degrees of freedom per space point. Moreover, using this symplectic framework, we yield the gauge transformations and the structure of the constraints directly from the zero-modes of the corresponding pre-symplectic matrix.

\end{abstract}

\maketitle
%%%%%%%%%%%%%%%%%%%% Introduction
\section{Introduction}

\label{INTRO}

Axions have received increased attention lately  as a natural candidate in the search for new physics effects, which may help us to shed light on the yet unanswered questions of contemporary physics.  In the context of high-energy particle physics, the axion is a dynamical scalar degree of freedom, which was first suggested as a solution to the strong CP problem via the Peccei-Quinn mechanism \cite{Peccei, Peccei2, Wilczek0,Wilczek, Weinberg, Jihn, Kim,Shifman,Dine,Zhitnitsky}. In addition to solving the strong CP problem, the axion turns out to be one of the most promising candidates in forming the dark matter that dominates the matter density of our Universe \cite{axion_dark, Bradley, Vergados,Camila, Lombardo}, as it is stable on cosmological time scales due to its tiny mass and  interacts very weakly  with Standard Model fields. In such a context, the axions may be produced non-thermally in the early universe, for example, through the misalignment mechanism \cite{Preskill,Nelson,Raymond}. Interestingly enough, the axion is not only a dark matter candidate and/or a solution to the strong CP problem, but it also provides a unique window through which one may explore a theoretical prediction of string theory. Such a theory predicts a plethora of ultra-light axions that  arise from the string compactification \cite{Arvanitaki}. So, in essence, one of the most remarkable features of the axions is the axion-photon mixing and oscillation in the presence of a strong magnetic field \cite{Raffelt}. More recently, in the context of condensed-matter physics, axionic degrees of freedom (DoF) are predicted to materialize as low-energy quasi-particles in certain materials such as topological insulators (TIs) \cite{Qi0,Rundong,Akihiko,Nenno,Katsuhisa,Xia,Zhang,DMEF,Tokura}; a new class of phases of matter which is of purely topological origin. Since these quasi-particles inside TIs have an interaction with the electromagnetic field, similar to the axions from particle physics, they are  usually called axions \cite{Rundong,Akihiko, Nenno,Katsuhisa}. However, compared with its  particle physics version, the mass of these quasi-axions is never zero. 

From an experimental point of view, it is worth commenting that the quasi-axions proposed in condensed-matter physics have the advantage of being able to be observed  inside TIs in controlled experimental settings \cite{Xia,Zhang, Wilczek2, DMEF}.  Whereas the dark matter axion has not yet  been detected. The most popular strategy to search for an axion signal experimentally is the axion haloscope technique suggested by Sikivie \cite{Sikivie, Sikivie1}, which is based on the resonant conversion of the axions into single microwave photons carrying its full energy in a cavity permeated by a strong magnetic field \cite{Stodolsky,Primakoff}. Accordingly, there are extensive ongoing and future experimental efforts to detect the dark matter axion by using the cavity haloscope technique (see e.g. ref. \cite{Igor}), such as ADMX \cite{ADMX}, HAYSTAC \cite{HAYSTAC}, CULTASK \cite{CULTASK}, MADMAX \cite{MADMAX}, ABRACADABRA \cite{ABRACADABRA}, and many other \cite{Orpheus, ORGAN, ORGAN2,RADES,David,Engel}

Along these lines, it is widely known that the axion develops generic couplings to photons,  whose dynamics can be generally described by adding a $\mathcal{CP}$ violating topological term of the form $\mathcal{L}_{\text{Axion}}\propto\theta \epsilon^{\mu\nu\alpha\beta}F_{\mu\nu}F_{\alpha\beta}$ to the Maxwell Lagrangian $\mathcal{L}_{\text{Maxwell}}\propto F^{\mu\nu}F_{\mu\nu}$, where $\theta$ is the axion field and $F_{\mu\nu}$ stands for the electromagnetic field strength tensor. The result is an effective theory that is usually referred to as axion electrodynamics, or more generally, modified axion electrodynamics.  In particular, when $\theta$ is \textit{static} the topological term only contributes with a surface term $\partial_{\mu}\mathcal{J}_{\text{C-S}}^{\mu}$ of the four-dimensional Chern-Simons current $\mathcal{J}^{\mu}_{\text{C-S}}=\epsilon^{\mu\nu\alpha\beta}A_{\nu}F_{\alpha\beta}$, which does not contribute to Maxwell's equations, as can be readily seen after using  the Bianchi identity $\epsilon^{\alpha\beta\mu\nu}\partial_{\beta}F_{\mu\nu} = 0$. This means that the axion contributions to Maxwell's equations arise only when $\theta$ is dynamical, i.e. when $\partial_{\alpha}\theta\neq0$.  To our knowledge, the $\mathcal{L}_{\text{Axion}}$ term can also be written up to a boundary term, as  $\mathcal{L}_{\text{Axion}}\propto\partial_{\mu}\theta \mathcal{J}^{\mu}_{\text{C-S}}$,  showing that $\theta$ is manifestly \textit{dynamical}. In this scenario, the  total derivative can be omitted as it does not alter the bulk dynamics of  axion electrodynamics.   Recently, in Ref. \cite {Raymond}, a new  misalignment mechanism with a nonzero initial velocity ($\partial_{0}{\theta}\neq0$) has been presented, where the dark matter abundance of a generic axion particle could be determined by the initial field velocity, as opposed to the conventionally assumed initial misalignment.

Given the considerable interest that  the axion field has  acquired in different areas of physics as a portal to search for signs of new physics, a meaningful description of the conserved quantities and the Hamiltonian dynamics may substantially enrich the physical content of modified axion electrodynamics. For instance, it is only in the Hamiltonian formalism that we can do  a reliable counting of the physical DoF in a given theory, and thermodynamic quantities such as energy, work, and entropy are naturally defined there \cite{Allahverdyan}. The Hamiltonian description of any physical theory consists of a phase space equipped with a symplectic structure  (or, equivalently, the Poisson bracket), and a Hamiltonian functional. The phase space is a specification of the DoF in the theory, while the Hamiltonian functional describes how these DoF evolve in time.  The split between the Hamiltonian functional and the phase space is also relevant when quantizing the system, as one must replace the phase space by a Hilbert space, while the Hamiltonian functional is replaced by a time evolution operator \cite{Josh,Henneaux}. On the other hand, every symmetry in the action formulation of particle or field theories implies a conservation law according to Noether's theorems \cite{Noether}. For instance, by virtue of Noether's first theorem, the invariance of the action under translation in time, translation in space, and rotation, implies the existence of the conservation of energy, linear momentum, and angular momentum, respectively.   Besides that, Noether's second theorem leads to a conservation law associated to local gauge symmetry as well.

Motivated by the above considerations, our goal in this work is thus to present a full and detailed analysis of conserved quantities and the Hamiltonian dynamics of modified axion electrodynamics. To do so, we shall apply the covariant phase space method and Faddeev-Jackiw's Hamiltonian formalism for constrained systems. The former approach provides a powerful tool for defining conserved charges in a covariant way for generic theories with local gauge symmetries \cite{Witten, Noether-Wald, Wald}. The procedure is carried out by keeping track of the surface terms in the variation of the action. Meanwhile, Faddeev-Jackiw's formalism allows us to compute the gauge transformations, the structure of the constraints, the correct number of DoF, the quantization brackets, and the functional measure on the path integral, by studying only the properties of the pre-symplectic matrix and its corresponding zero-modes. \cite{F-J, Wotzasek, Montani, Toms}. The advantage of the symplectic approach is that we do not need to classify the constraints into first- and second-class ones, and therefore it is different from Dirac's approach \cite{Henneaux, Dirac}. Furthermore, we shall use the Belinfante-Rosenfeld procedure to determine the covariant improved energy-momentum tensor for the interacting theory \cite{ Belinfante2, Rosenfeld, Daniel}.

The present work is organized as follows. We begin Section \ref{theory} by defining the action principle corresponding to modified axion electrodynamics. Subsequently, we derive the equations of motion and the boundary term. Section \ref{charges} presents the construction of conserved charges associated with the gauge symmetry as well as the canonical and improved energy-momentum tensor. In Section \ref{canonical}, we obtain both the constraint structure and the gauge transformations by using the zero-modes of the pre-symplectic matrix. We then determine the fundamental F-J brackets and the functional measure on the path integral associated with our model by introducing a gauge-fixing procedure. Finally, we present the conclusions in Section. \ref{final}.

\section{The action principle}
\label{theory}
Since the axion is  a pseudo-Nambu-Goldstone boson arising from the spontaneous breakndown of global Peccei-Quinn symmetry $U(1)_\text{PQ}$ \cite{Peccei, Peccei2, Wilczek, Weinberg, Jihn},  it can generically develop couplings to  Standard Model particles, e.g. photons, through its  derivative, originated from a universal coupling,
\begin{equation}
\sim\mathcal{J}_{\text{PQ}}^{\mu}\partial_{\mu}\theta,
\end{equation}
where $\mathcal{J}_{\text{PQ}}^{\mu}= \bar{\Psi}\gamma^{\mu}\gamma^{5}\Psi$ is the current of the global  chiral $U(1)_{\text{PQ}}$ symmetry \cite{PDG}, yielding the abelian version of the chiral anomaly $\partial_{\mu}\mathcal{J}_{PQ}^{\mu}\propto\epsilon^{\mu\nu\alpha\beta}F_{\mu\nu}F_{\alpha\beta}$. Although the strength of this coupling is very small for the axion-photon coupling, a method to detect axion dark matter using this type of coupling inside a TI has been recently proposed in Ref. \cite{David, Engel}. In this work, specifically, we consider the effective action principle for modified axion electrodynamics, describing the generic coupling of massless axions to photons, defined by  \cite{Wilczek,Carrol,Dunne,Dmitri, Tobar}
\begin{equation}
S[A,\theta]=\int_{\mathcal{M}}d^{4}x\left(-\frac{1}{16\pi}F_{\mu\nu}F^{\mu\nu}+\frac{\kappa}{16\pi^{2}}\widetilde{F}^{\mu\nu}A_{\left[\mu\right.}\partial_{\left.\nu\right]}\theta +\frac{1}{2}\partial_{\mu}\theta\partial^{\mu}\theta-J^{\mu}A_{\mu}\right).\label{action}
\end{equation}
Here $\mathcal{M}=\mathbb{R}\times\Sigma$ is a four-dimensional space-time, where $\Sigma$ represents the three-dimensional space-like hypersurface and $\mathbb{R}$ is a time interval, $\kappa$  is the parameter characterizing  the strength of the axion-photon coupling.  $\theta(x)$  is the axion field, $\widetilde{F}^{\mu\nu}=(1/2)\epsilon^{\mu\nu\beta\gamma}F_{\beta\gamma}$, and $J^{\mu}=(\varrho,\mathbf{J})$ stands for the electric 4-current. In the following, we use the signature $(1,-1,-1,-1)$ for the metric, and we further denote the components of $x$  as $x^{\mu}$, $\mu=0,1,2,3$  and those of $\mathbf{x}$ as $x^{a}$, $a=1,2,3$.

Let us begin by deriving the field equations for modified axion electrodynamics. An arbitrary variation of the action (\ref{action}) gives the field equations plus a boundary term:
\begin{eqnarray}
\delta S[A_{\mu},\theta]&=&\int_{\mathcal{M}}d^{4}x\left[\left(\frac{1}{4\pi}\partial_{\mu}F^{\mu\nu}-\frac{\kappa}{4\pi^{2}}\widetilde{F}^{\mu\nu}\partial_{\mu}\theta-J^{\nu}\right)\delta A_{\nu}-\partial_{\mu}\left(\partial^{\mu}\theta-\frac{\kappa}{8\pi^{2}}\widetilde{F}^{\mu\nu}A_{\nu}\right)\delta\theta\right.\nonumber\\
&&+\left.\partial_{\mu}\left(\left(\partial^{\mu}\theta-\frac{\kappa}{8\pi^{2}}\widetilde{F}^{\mu\nu}A_{\nu}\right)\delta\theta-\frac{1}{4\pi}\left(F^{\mu\nu}-\frac{\kappa}{2\pi}\epsilon^{\mu\nu\alpha\beta}A_{\alpha}\partial_{\beta}\theta\right)\delta A_{\nu}\right)\right].\label{variation}
\end{eqnarray}
As we can see, the first term simply contains the following  bulk equations of motion
\begin{equation}
\delta A^{\nu}:\partial_{\mu}F^{\mu\nu}=\frac{\kappa}{\pi}\widetilde{F}^{\mu\nu}\partial_{\mu}\theta+4\pi J^{\nu},\label{Ec-mov-1}
\end{equation}
here the first term on the right-hand side of  Eq. (\ref{Ec-mov-1}) represents the additional contribution to Gauss's and Amp\`{e}re's laws respectively, due to the presence of the dynamical axion field, which leads to an additional current density: $J_{\theta}^{\nu}=\partial_{\mu}\left((\kappa/\pi)\theta\widetilde{F}^{\mu\nu}\right)=(\kappa/\pi)\widetilde{F}^{\mu\nu}\partial_{\mu}\theta$. In other words, the spatial gradient and time-dependence of $\theta$ in an external electromagnetic field acts as a new source for the standard Maxwell's equations. Physically, such a topological current describes the anomalous Hall effect \cite{Naoto} (induced by the spatial gradient of $\theta$) and the dynamical chiral magnetic effect (generated by magnetic fields from the temporal gradient of $\theta$) \cite{Dunne}. Consequently, taking into account the Lorenz gauge defined by $\partial_{\mu}A^{\mu}=0$, the equation of motion for $A^{\mu}$ (\ref{Ec-mov-1}) reduce to
\begin{equation}
\left(-\eta^{\nu}_{\beta}\Box-\frac{\kappa}{\pi}\partial_{\mu}\theta\epsilon^{\mu\nu\alpha}{_{\beta}}\partial_{\alpha}\right)A^{\beta}=4\pi J^{\nu},
\end{equation}
where $\Box\equiv\partial_{\mu}\partial^{\mu}$ is the d'Alembertian. A general solution of these coupled differential equation can be expressed by using the Green's function \cite{Alberto, OmarJ}.

On the other hand, it is evident that the variation of the action (\ref{action}) does not give rise to the full set of Maxwell equations. Nevertheless, the homogeneous axion electrodynamics equations that express the field-potential relationship, i.e., Gauss law for magnetism and Faraday law, can be derived from a geometric property of electrodynamics, i.e. the Bianchi identity,
\begin{equation}
\partial_{\mu}\widetilde{F}^{\mu\nu}= 0,
\end{equation}
which does not get modified by including the axion field.

Moreover, from the second term of Eq. (\ref{variation}), one gets the massless axion field equation
\begin{equation}
\delta\theta:\partial_{\mu}\left(\partial^{\mu}\theta-\frac{\kappa}{8\pi^{2}}\tilde{F}^{\mu\nu}A_{\nu}\right)=0,
\end{equation}
which implies the existence of a conserved topological current in the sense that $\partial_{\mu}\widetilde{\mathcal{J}}^{\mu}=0$, where $\widetilde{\mathcal{J}}^{\mu}$ is given by
\begin{equation}
\widetilde{\mathcal{J}}^{\mu}=\partial^{\mu}\theta-\frac{\kappa}{8\pi^{2}}\left(\epsilon^{\mu\nu\alpha\beta}A_{\nu}\partial_{\alpha}A_{\beta}\right).\label{C-S}
\end{equation}
From the above equation it can be easily seen that the second term is the Chern-Simons term  for photon, which instead of being defined in three-dimensions it is defined in four-dimensions. Hence, the axion's influence can be interpreted as a topological effect.

Finally, it is worth noticing that to have a well-defined action principle, it is necessary that field equations hold and that the surface term vanishes for a set of given boundary conditions on fields, that is
\begin{eqnarray}
\left.\left(\partial^{\mu}\theta-\frac{\kappa}{8\pi^{2}}\epsilon^{\mu\nu\alpha\beta}A_{\nu}\partial_{\alpha}A_{\beta}\right)\right\arrowvert_{\partial\mathcal{M}}&=&0,\\
\left.\left(F^{\mu\nu}-\frac{\kappa}{2\pi}\epsilon^{\mu\nu\alpha\beta}A_{\alpha}\partial_{\beta}\theta\right)\right\arrowvert_{\partial\mathcal{M}}&=&0.
\end{eqnarray}

At this point, it is important to remark   that all the above set of field equations derived from  Eq. (\ref{action}) constitute the axion electrodynamics theory including the dynamical axion field.

\section{Conserved charges and energy-momentum tensor}
\label{charges}
For simplicity, let us consider modified axion electrodynamics with no external sources, i.e.  $J^{\mu}=0$. To compute the charges we can use the covariant phase space method \cite{Noether-Wald,Wald}. According to the action principle, when we calculate the variation of the action and restrict to on-shell field configurations, the result is a total derivative which defines the pre-symplectic potential density $\Theta^{\mu}$ for the theory on the space of solutions to the field equations. In our case, we immediately see that
\begin{equation}
\Theta^{\mu}\left[\Xi_{I},\delta\Xi_{I}\right]=\left(\partial^{\mu}\theta-\frac{\kappa}{8\pi^{2}}\widetilde{F}^{\mu\nu}A_{\nu}\right)\delta\theta-\frac{1}{4\pi}\left(F^{\mu\nu}-\frac{\kappa}{2\pi}\epsilon^{\mu\nu\alpha\beta}A_{\alpha}\partial_{\beta}\theta\right)\delta A_{\nu}.
\end{equation}
Here all dynamical fields in the theory are collectively denoted by $\Xi_{I}$, and therefore $\delta\Xi_{I}$ is a  generic set of  field perturbations. Now, given two independent field variations $\delta_{1}\Xi_{I}$ and $\delta_{2}\Xi_{I}$, the antisymmetrized variation of the pre-symplectic potential density $\Theta^{\mu}$ defines  the pre-symplectic  current  $\omega^{\mu}\left[\Xi_{I},\delta_{1}\Xi_{I},\delta_{2}\Xi_{I}\right]\equiv\delta_{1}\Theta^{\mu}\left[\Xi_{I},\delta_{2}\Xi_{I}\right]-\delta_{2}\Theta^{\mu}\left[\Xi_{I},\delta_{1}\Xi_{I}\right]$. Accordingly, we find
\begin{eqnarray}
\omega^{\mu}\left[\Xi_{I},\delta_{1}\Xi_{I},\delta_{2}\Xi_{I}\right]&=&\left(\partial^{\mu}\delta_{1}\theta-\frac{\kappa}{8\pi^{2}}A_{\nu}\delta_{1}\widetilde{F}^{\mu\nu}-\frac{\kappa}{8\pi^{2}}\widetilde{F}^{\mu\nu}\delta_{1}A_{\nu}\right)\delta_{2}\theta\nonumber\\
&&-\frac{1}{4\pi}\left(\delta_{1}F^{\mu\nu}-\frac{\kappa}{2\pi}\epsilon^{\mu\nu\alpha\beta}\left(A_{\alpha}\partial_{\beta}\delta_{1}\theta+\partial_{\beta}\theta\delta_{1}A_{\alpha}\right)\right)\delta_{2} A_{\nu}-\left(1\leftrightarrow2\right).\label{Current}
\end{eqnarray}
To define the conserved charges, we need to compute $\omega^{\mu}\left[\Xi_{I},\delta\Xi_{I},\delta_{\eta}\Xi_{I}\right]$ with $\delta_{\eta}$ being a specific set of transformations parametrized by an arbitrary scalar function $\eta$. Using the gauge transformations for modified axion electrodynamics (\ref{gauge1}) and (\ref{gauge2}) (we will derive  these transformations below after discussing the constraints structure of the theory in Section \ref{canonical}),  Eq. (\ref{Current})  takes the form
\begin{equation}
\omega^{\mu}\left[\Xi_{I},\delta\Xi_{I},\delta_{\eta}\Xi_{I}\right]=\frac{1}{4\pi}\left[\delta\left(-F^{\mu\nu}+\frac{\kappa}{\pi}\theta\tilde{F}^{\mu\nu}\right)+\frac{\kappa}{2\pi}\epsilon^{\mu\nu\alpha\beta}\partial_{\beta}\left(2\theta \delta A_{\alpha}+A_{\alpha}\delta\theta\right)\right]\partial_{\nu}\eta. \label{current0}
\end{equation}
Using the  equations of motion for the linearized field perturbations,  the pre-symplectic current   (\ref{current0}) can be written, on-shell, as
\begin{equation}
\omega^{\mu}\left[\Xi_{I},\delta\Xi_{I},\delta_{\eta}\Xi_{I}\right]=\frac{1}{4\pi}\partial_{\nu}\left[\delta\left( -F^{\mu\nu}+\frac{\kappa}{\pi}\theta\tilde{F}^{\mu\nu}\right)\eta-\frac{\kappa}{2\pi}\epsilon^{\mu\nu\alpha\beta}\partial_{\alpha}\left( 2\theta\delta A_{\beta}+A_{\beta}\delta\theta\right)\eta\right]. \label{current}
\end{equation}
Hence, the pre-symplectic current $\omega^{\mu}$ contracted with the gauge transformations $\delta_{\eta}$ is an exact on-shell form. Notwithstanding the above, the pre-symplectic current $\omega^{\mu}$  is ill-defined due to the presence of the last term; however,  it is important to note that $\Theta$ can be shifted as follow \cite{Noether-Wald, Wald}
\begin{eqnarray}
\Theta'^{\mu}\left[\Xi_{I},\delta\Xi_{I}\right]\longrightarrow\Theta^{\mu}\left[\Xi_{I},\delta\Xi_{I}\right] + \delta\Delta^{\mu}\left[\Xi_{I}\right],
\end{eqnarray}
such that, once we have taken the contribution of $\Delta$ into account, the ambiguity in Eq. (\ref{current}) is removed. Choosing $\Delta^{\mu}=(\kappa/8\pi^{2})\theta\epsilon^{\mu\nu\alpha\beta}A_{\nu}\partial_{\alpha}A_{\beta}$ yield the new pre-symplectic current
\begin{equation}
\omega'^{\mu}\left[\Xi_{I},\delta\Xi_{I},\delta_{\eta}\Xi_{I}\right]=\frac{1}{4\pi}\delta\left( - F^{\mu\nu}+\frac{\kappa}{\pi}\theta\tilde{F}^{\mu\nu}\right)\partial_{\nu}\eta,
\end{equation}
which can be rewritten as
\begin{equation}
\omega'^{\mu}\left[\Xi_{I},\delta\Xi_{I},\delta_{\eta}\Xi_{I}\right]=\frac{1}{4\pi}\partial_{\nu}\left[\delta\left(F^{\mu\nu}-\frac{\kappa}{\pi}\theta\tilde{F}^{\mu\nu}\right)\right]\eta+\frac{1}{4\pi}\partial_{\nu}\left[\delta\left( - F^{\mu\nu}+\frac{\kappa}{\pi}\theta\tilde{F}^{\mu\nu}\right)\eta \right]. \label{current2}
\end{equation}
It has been assumed that the boundary $\partial\mathcal{M}$ of  $\mathcal{M}$ admits a canonical split with the identification of a Cauchy hypersurface $\Sigma$, and that $\Sigma$  has a boundary $\partial\Sigma$ a 2-surface in $\partial\mathcal{M}$, which is often the case in situations of physical relevance. It guarantees that the phase space contains all degrees of freedom. We can then  integrate $\omega'$ over the Cauchy surface to define on-shell the charge  perturbations $\delta\mathcal{Q}'_{\eta}$ associated with $\eta$,
\begin{equation}
\delta\mathcal{Q}'_{\eta}=\int_{\Sigma}\omega'\left[\Xi_{I},\delta\Xi_{I},\delta_{\eta}\Xi_{I}\right]=\frac{1}{4\pi}\oint_{\partial\Sigma}d^{2}\Sigma_{\mu\nu}\delta\left(- F^{\mu\nu}+\frac{\kappa}{\pi}\theta(x)\tilde{F}^{\mu\nu}\right)\eta. \label{current2}
\end{equation}
From this it is clear that the pre-symplectic structure defined by $\Omega\equiv\int_{\Sigma}\omega'$ has degenerate gauge directions inside the Cauchy surface, but not on its boundary. Therefore, the Noether charge is given by a codimension-two integral, it simply means that the conserved charge is a surface charge. In this way, by integrating over variations, we  find out that the finite charge of the system turns out to be
\begin{equation}
\mathcal{Q}'_{\eta}=\oint_{\partial\Sigma}d^{2}\Sigma_{\mu\nu}\left(-\frac{1}{4\pi}F^{\mu\nu}+\frac{\kappa}{4\pi^{2}}\theta(x)\widetilde{F}^{\mu\nu}\right)\eta.\label{Chargefinal}
\end{equation}
Besides, it follows that in the scenario where $\eta\longrightarrow 1$ near the boundary, we find that $\mathcal{Q}'=\oint_{\partial\Sigma}d^{2}\Sigma_{\mu\nu}\left(-(1/4\pi)F^{\mu\nu}+(\kappa/4\pi^{2})\theta(x)\widetilde{F}^{\mu\nu}\right)$, which simply reduces to the Noether electric charge plus a \textit{magnetic} charge induced by the space-time dependent axion coming from a topological origin, as shown in Eq. (\ref{C-S}).

Furthermore, the charges can be written in terms of electric and magnetic fields $\mathbf{E}$ and $\mathbf{B}$. This can be achieved by taking the Cauchy surface $\Sigma$ to be the $t=\text{constant surface}$, such that the boundary $\partial\Sigma_{t}$ is a sphere $S$ of constant radius $r\rightarrow\infty$. Then we can use the standard notation $\mathbf{E}=E^{a}= F^{a0}$, $\mathbf{B}=B^{a}= \epsilon^{0abc}\partial_{b}A_{c}$, for the electric and magnetic field intensity, respectively. Hence from our previously calculated expression for $\mathcal{Q}'$, Eq. (\ref{Chargefinal}), we find that
\begin{equation}
\mathcal{Q}'=\frac{1}{4\pi}\oint_{S}d^{2}S \left(\mathbf{E}+\frac{\kappa}{\pi}\theta(x)\mathbf{B}\right)\cdot\mathbf{n}.
\end{equation}
where $\mathbf{n}$ is a unit vector normal to the sphere. This shows that the conserved charge is the sum of the Noether electric and magnetic charge induced by the dynamical axion field. At this point, it is worth noting that a composite particle with both electric and magnetic charges, would be what in elementary particle physics is called a dyon \cite{witten}.

On the other hand, the invariance of the action functional given by (\ref{action}) under space-time translations, implies the local conservation law for the canonical energy-momentum tensor, i.e. $\partial_{\mu}T^{\mu\nu}_{\text{can}}=0$. Such a canonical stress tensor, in the absence of a source of photons, can be obtained from the Lagrangian density in Eq. (\ref{action}) in a straightforward way \cite{Noether}:
\begin{eqnarray}
T^{\mu\nu}_{\text{can}}&\equiv &\frac{\partial\mathcal{L}}{\partial(\partial_{\mu}A_{\gamma})}\partial^{\nu}A_{\gamma}+\frac{\partial\mathcal{L}}{\partial(\partial_{\mu}\theta)}\partial^{\nu}\theta-\eta^{\mu\nu}\mathcal{L}\\
&=&-\frac{1}{4\pi}F^{\mu\alpha}\partial^{\nu}A_{\alpha}-\frac{\kappa}{8\pi^{2}}\epsilon^{\mu\alpha\beta\gamma}\partial_{\alpha}\theta A_{\beta}\partial^{\nu}A_{\gamma}-\frac{\kappa}{8\pi^{2}}\tilde{F}^{\mu\alpha}A_{\alpha}\partial^{\nu}\theta+\partial^{\mu}\theta\partial^{\nu}\theta-\eta^{\mu\nu}\mathcal{L},
\end{eqnarray}
which is neither symmetric in its indices nor gauge invariant in general and thereby needs to be improved.  This can be achieved by using the Belinfante-Rosenfeld procedure, whose goal is to construct out a symmetric energy-momentum tensor (EMT)  from the canonical stress tensor \cite{Belinfante2,Rosenfeld,Daniel}. In the Belinfante-Rosenfeld construction, the resulting improved EMT $T^{\mu\nu}_{\text{imp}}$ is given by
\begin{equation}
T^{\mu\nu}_{\text{imp}}\longrightarrow T^{\mu\nu}_{\text{can}}+\partial_{\gamma}\Sigma^{\mu\gamma\nu}.
\end{equation}
Here $\Sigma^{\mu\gamma\nu}=-\Sigma^{\gamma\mu\nu}$ is some superpotential antisymmetric in the last pair of indices, so that it is off-shell divergenceless, and therefore it does not spoil the conservation law, $\partial_{\nu}T^{\mu\nu}_{\text{imp}}=\partial_{\nu}T^{\mu\nu}_{\text{can}}+\partial_{\nu}\partial_{\gamma}\Sigma^{\mu\nu\gamma}=\partial_{\nu}T^{\mu\nu}_{\text{can}}=0$.  Choosing $\Sigma^{\mu\gamma\nu}=(1/4\pi)(F^{\mu\gamma}+(\kappa/2\pi)\partial_{\alpha}\theta\epsilon^{\alpha\beta\mu\gamma}A_{\beta})A^{\nu}$ yields the improved on-shell EMT  that describes the interaction of the axion field with the electromagnetic field
\begin{equation}
T^{\mu\nu}_{\text{imp}}=-\frac{1}{4\pi}F^{\mu\alpha}F^{\nu}{_{\alpha}}+\eta^{\mu\nu}\frac{1}{16\pi}F_{\alpha\beta}F^{\alpha\beta}+\frac{\kappa}{8\pi^{2}}A_{\alpha}\widetilde{F}^{\alpha\mu}\partial^{\nu}\theta+\partial^{\mu}\theta\partial^{\nu}\theta-\eta^{\mu\nu}\frac{1}{2}\partial_{\alpha}\theta\partial^{\alpha}\theta.\label{Belinfante}
\end{equation}
It is important to note that the term $(\kappa/8\pi^{2}) A_{\alpha}\widetilde{F}^{\alpha\mu}\partial_{\mu}\theta$  present in the action (\ref{action}), which contributes the third and sixth term to (\ref{Belinfante}), renders the improved non-symmetric energy-momentum tensor.  Nevertheless, $T^{\mu\nu}_{\text{imp}}$ satisfies the correct conservation law in the sense that its on-shell divergence vanishes, $\partial_{\mu}T^{\mu\nu}_{\text{imp}}=0$. Besides,  the above improved tensor (\ref{Belinfante}) reduces to Maxwell's stress tensor if we set $\theta=\text{constant}$. Remarkably, the  components of the improved EMT (\ref{Belinfante}) contain the proper energy density $\mathcal{E}$ ($T_{\text{imp}}^{00}$), the momentum density coincident with the total energy flux $\mathcal{P}^{a}$ ($T_{\text{imp}}^{0a}$), and also the energy flux $\mathcal{S}^{a}$ ($T_{\text{imp}}^{a0}$). In terms of fields and potentials they are given by
\begin{eqnarray}
\mathcal{E}&=&\frac{1}{8\pi}\left(\mathbf{E}^{2}+\mathbf{B}^{2}\right)+\frac{\kappa}{8\pi^{2}}\partial^{0}\theta\mathbf{B}\cdot\mathbf{A}+\frac{1}{2}\partial_{0}\theta\partial^{0}\theta-\frac{1}{2}\partial_{a}\theta\partial^{a}\theta,\label{E}\\
\mathcal{P}^{a}&=&\frac{1}{4\pi}\left(\mathbf{E}\times\mathbf{B}\right)^{a}+\frac{\kappa}{8\pi^{2}}\partial^{a}\theta\mathbf{B}\cdot\mathbf{A}+\partial^{0}\theta\partial^{a}\theta,\label{P}\\
\mathcal{S}^{a}&=&\frac{1}{4\pi}\left(\mathbf{E}\times\mathbf{B}\right)^{a}-\frac{\kappa}{8\pi^{2}}\partial^{0}\theta\left(\mathbf{A}\times\mathbf{E}\right)^{a}+\frac{\kappa}{8\pi^{2}}\partial^{0}\theta{B^{a}}A_{0}-\partial^{0}\theta\partial^{a}\theta.\label{S}
\end{eqnarray}
The second term in the components of the improved energy-momentum tensor  (\ref{E})-(\ref{P}) is identified as the magnetic helicity density $\mathbf{B}\cdot\mathbf{A}$, which measures the winding of magnetic lines of force and/or fluid vortex lines, whereas the second term in Eq. (\ref{S}) is the spin angular momentum density $\mathbf{A}\times\mathbf{E}$ of the  electromagnetic field. Let us finally point out that the helicity density is induced by the time derivative of the axion field which could also be identified with the chiral chemical potential $\mu_{5}=\partial_{0}\theta$ \cite{Dunne}.

\section{Canonical analysis in the symplectic framework}
\label{canonical}
We now proceed to the canonical analysis of the modified axion electrodynamics theory coupled to an external source $J^{0}$. We will apply the symplectic framework for constrained systems \cite{F-J, Wotzasek,Montani}. This formalism was introduced by Faddeev and Jackiw \cite{F-J}, to avoid consistency problems that spoil the Poisson brackets algebra and consequently fail any quantization techniques. Such a framework is geometrically well motivated and is based on the symplectic structure of the space phase. As shown by Faddeev and Jackiw, the constraints emerge as algebraic relations necessary to maintain the consistency of the equations of motion in systems with constraints. The advantage of this framework is that we do not need to catalog the constraints as first- and second-class ones like in Dirac's theory. In this setting, all the constraints are treated at the same footing. Moreover, it does not rely on Dirac's conjecture. Therefore it is a different method from the well-known Dirac approach \cite{Henneaux, Dirac}.  Still, several essential elements of a physical theory, such as the structure of all physical constraints, the local gauge symmetry and its generators, the quantization brackets structure, the functional measure for determining the quantum transition amplitude, and the correct number of physical degrees of freedom, can be systematically addressed by studying only the properties of the pre-symplectic matrix and its corresponding zero-modes. Once the classical field theory is obtained, quantization can proceed via a path integral or in a canonical way. Since symplectic formalism is not the most widely used, we will discuss it in more detail below.

At first, we begin by slicing the four-dimensional spacetime manifold, separating the temporal components from the spatial ones of our fields and their derivatives without fixing any gauge. After performing the $(3+1)$ decomposition, the action (\ref{action}) can be expressed as
\begin{eqnarray}
S&=&\int_\mathbb{R} dt L=\int_\mathbb{R} dt \int_{\Sigma} d\mathbf{x}\left[A_{0}\left(\frac{1}{8\pi}\partial_{a}F^{a0}+\frac{\kappa}{8\pi^{2}}\epsilon^{0abc}P_{a}F_{bc}-J^{0}\right)+\dot{A}_{a}\left(\frac{1}{8\pi}F^{a0}+\frac{\kappa}{8\pi^{2}}\epsilon^{0abc}A_{b}P_{c}\right)\right.\nonumber\\
&&+\left.\dot{\theta}\left(\frac{1}{2}\dot{\theta}-\frac{\kappa}{16\pi^{2}}\epsilon^{0abc}A_{a}F_{bc}\right)-\frac{1}{16\pi}F_{ab}F^{ab}+\frac{1}{2}P_{a}P^{a}+\partial_{a}\left(\frac{1}{8\pi}A_{0}\left(F^{0a}-\frac{\kappa}{\pi}\epsilon^{0abc}A_{b}P_{c}\right)\right)\right].\label{3+1}
\end{eqnarray}
Here $F_{ab}=\partial_{a}A_{b}-\partial_{b}A_{a}$ is the field strength of $A_{a}$ and $P_{a}=\partial_{a}\theta$.
To proceed with the canonical analysis, it is convenient to transform the Lagrangian in Eq. (\ref{3+1}) into a first-order Lagrangian. To this end, we define the canonical momenta $(\Pi^{0},\Pi^{a},P)$ conjugate to the fields $(A_{0}, A_{a},\theta)$, respectively, as
\begin{eqnarray}
\Pi^{0}&\equiv&\frac{\delta{L}}{\delta\dot{A}_{0}}=0,\label{mom1}\\
\Pi^{a}&\equiv&\frac{\delta{L}}{\delta\dot{A}_{a}}=\frac{1}{4\pi}\left(F^{a0}+\frac{\kappa}{2\pi}\epsilon^{0abc}A_{b}P_{c}\right),\label{mom2}\\
P&\equiv&\frac{\delta{L}}{\delta\dot{\theta}}=\dot{\theta}-\frac{\kappa}{16\pi^{2}}\epsilon^{0abc}A_{a}F_{bc}.\label{mom3}
\end{eqnarray}
Thus, after performing a Legendre transformation, we can rewrite the Lagrangian in Eq.  (\ref{3+1}) as a first-order one in time derivatives as follows
\begin{eqnarray}
L&=&\int_{\Sigma}d\mathbf{x}\left[\dot{A}_{a}\Pi^{a}+\dot{\theta}P+A_{0}\left(\partial_{a}\Pi^{a}+\frac{\kappa}{16\pi^{2}}\epsilon^{0abc}F_{ab}P_{c}-J^{0}\right)-\frac{1}{16\pi}F_{ab}F^{ab}+\frac{1}{2}P_{a}P^{a}\right.\nonumber\\
&&+\left.2\pi\left(\Pi^{a}-\frac{\kappa}{8\pi^{2}}\epsilon^{0abc}A_{b}P_{c}\right)^{2}-\frac{1}{2}\left(P+\frac{\kappa}{16\pi^{2}}\epsilon^{0abc}A_{a}F_{bc}\right)^{2}-\partial_{a}\left(A_{0}\Pi^{a}\right)\right],\label{Lagran-first-order}
\end{eqnarray}
which is convenient for analyzing the dynamics on $\Sigma$ in the symplectic framework. Thus, without loss of generality, the action functional Eq. (\ref{Lagran-first-order}) can be schematically written in the symplectic form, up to a boundary term, as
\begin{equation}
S=\int_\mathbb{R} dt \int_{\Sigma} \left(a_{I}(\xi)\dot{\xi}^{I}-\mathcal{H}(\xi)\right)d\mathbf{x},\label{Lag_Sym}
\end{equation}
which is first-order in time derivative $\dot{\xi}$, where $\xi$ stands for the collection of all the fields and momenta of the theory. The first term in Eq. (\ref{Lag_Sym}) defines the so-called canonical one-form $a=a_{I}(\xi)\xi^{I}$, whereas the second term represents the symplectic potential which can also be identified with the canonical Hamiltonian density $\mathcal{H}$. For axion electrodynamics, the set of initial symplectic variables as well as the corresponding canonical one-form can be identified easily from the first-order Lagrangian (\ref{Lagran-first-order}) as follows
\begin{eqnarray}
{\xi}{^{I}}&=& (A{_{0}},A{_{a}},\Pi^{a}, \theta,P), \label{variables1}\\
{a}{_{I}} &=& ( 0,\Pi^{a},0, P,0 ).\label{form1}
\end{eqnarray}
On the other hand, the Hamiltonian density can be read off from the Lagrangian (\ref{Lagran-first-order})
\begin{eqnarray}
\mathcal{H}&=&-A_{0}\left(\partial_{a}\Pi^{a}+\frac{\kappa}{16\pi^{2}}\epsilon^{0abc}F_{ab}P_{c}-J^{0}\right)+\frac{1}{16\pi}F_{ab}F^{ab}-\frac{1}{2}P_{a}P^{a}\nonumber\\
&&-2\pi\left(\Pi^{a}-\frac{\kappa}{8\pi^{2}}\epsilon^{0abc}A_{b}P_{c}\right)^{2}+\frac{1}{2}\left(P+\frac{\kappa}{16\pi^{2}}\epsilon^{0abc}A_{a}F_{bc}\right)^{2}.
\end{eqnarray}
In the symplectic picture, the equations of motion deduced from the above action principle (\ref{Lag_Sym}) can be written  in a first-order form as
\begin{equation}
\int d\mathbf{x}\left(\mathcal{F}_{IJ}(\mathbf{x},\mathbf{y})\dot{\xi}(\mathbf{x})^{J}-\frac{\delta \mathcal{H}(\mathbf{x})}{\delta\xi^{I}(\mathbf{y})}\right)=0.\label{eqmot}
\end{equation}
Note that the dynamics of the model is then characterized by the so-called pre-symplectic two-form matrix, which is defined by
\begin{equation}
\mathcal{F}_{IJ}(\mathbf{x},\mathbf{y})\equiv\frac{\delta a_{J}(\mathbf{x})}{\delta\xi^{I}(\mathbf{y})}-\frac{\delta a_{I}(\mathbf{y})}{\delta\xi^{J}(\mathbf{x})}.\label{symplectic_matrix}
\end{equation}
Clearly $\mathcal{F}_{IJ}$ is an antisymmetrical matrix that can be either singular or non-singular. According to the symplectic approach, if the matrix $\mathcal{F}_{IJ}$ is non-singular, then its inverse can be obtained by solving the following relation:
\begin{equation}
\int d\mathbf{z}\left(\mathcal{F}_{IK}(\mathbf{x},\mathbf{z})\right)^{-1}\mathcal{F}^{KJ}(\mathbf{z},\mathbf{y})=\delta_{I}^{J}\delta^{3}(\mathbf{x}-\mathbf{y}),
\end{equation}
where $\left(\mathcal{F}_{IK}\right)^{-1}$ denotes the inverse matrix of $\mathcal{F}^{KJ}$. As a  consequence, the Eqs. (\ref{eqmot}) can be solved in terms of $\dot{\xi}_{I}$:
\begin{equation}
\dot{\xi}_{I}(\mathbf{y})=\int d\mathbf{z}(\mathcal{F}^{IJ}(\mathbf{y},\mathbf{z}))^{-1}\int d\mathbf{x}\frac{\delta \mathcal{H}(\mathbf{x})}{\delta\xi^{I}(\mathbf{z})}.
\end{equation}
In our model, the corresponding pre-symplectic matrix, using expression (\ref{symplectic_matrix}) and symplectic variable set (\ref{variables1}) as well as corresponding one-form (\ref{form1}),  is explicitly given as follows,
\begin{eqnarray}
\label{matrx_sym}\mathcal{F}_{IJ}(\mathbf{x},\mathbf{y})=
\left(
  \begin{array}{cccccc}
 0      &  0   &  0     &  0  &  0 	 	 \\
0    &  0      &  -\delta^{a}_{b}   &   0   &  0 	\\
    0      &  \delta^{a}_{b}    & 0  & 0	 &  0 \\
0  &   0  &0  & 0 	& -1 	 \\
0   &  0  & 0   &  1 	&  0
 \end{array}
\right) \delta^{3}(\mathbf{x}-\mathbf{y}).\label{f1}
\end{eqnarray}
It is not difficult to see that the two-form matrix (\ref{f1}) is certainly singular, since $\text{det} |\mathcal{F}_{IJ} |=0$, which implies that the theory is endowed with constraints. In other words, $\mathcal{F}_{IJ}$ is degenerate which means that there are more degrees of freedom in the equations of motion (\ref{eqmot}) than physical degrees of freedom in the theory. This suggests the existence of constraints in the system which must remove the unphysical degrees of freedom. In the symplectic formalism, the constraints emerge as algebraic relations necessary to maintain the consistency of the equations of motion. It is straightforward to determine that $\mathcal{F}_{IJ}$ has the following zero-mode:
\begin{equation}
v^{I}= \left (v^{A{_{0}}},0 ,0,0,0\right),\label{mode1}
\end{equation}
where $v^{A{_{0}}}$ is a totally arbitrary function. According to the symplectic formalism, the multiplication of this zero-mode by the gradient of the Hamiltonian leads to the following expression:
\begin{equation}
\int d\mathbf{y}\, v^{I}(\mathbf{y})\int \ d\mathbf{x}\frac{\delta\mathcal{H}(\mathbf{x})}{\delta\xi^{I}(\mathbf{y})}  =\int d\mathbf{y}\,v^{A{_{0}}}\left(\partial_{a}\Pi^{a}+\frac{\kappa}{16\pi^{2}}\epsilon^{0abc}F_{ab}P_{c}-J^{0}\right)=0,\label{p1}
\end{equation}
since $v^{A{_{0}}}$ is an arbitrary function, we obtain the following constraint on the dynamics:
\begin{equation}
\Omega=\partial_{a}\Pi^{a}+\frac{\kappa}{16\pi^{2}}\epsilon^{0abc}F_{ab}P_{c}-J^{0}=0,\label{fj_constraint}
\end{equation}
this makes evident that the zero-modes of the pre-symplectic matrix encode the information of the constraint. This constraint is the Gauss-law constraint reflecting the gauge invariance of the theory. Now, there is a natual consistency condition requeiring the stablility of the constraints surface under the time evolution given by $\dot{\Omega}=0$. At this point, it is worthwhile noting that the consistency condition, $\dot{\Omega}=\int d\mathbf{x}\left[\delta\Omega(\mathbf{x})/\delta\xi_{I}(\mathbf{y})\right]\dot{\xi}^{I}(\mathbf{y})=0$, depend only on the time derivative of the symplectic variables $\xi^{I}$. Therefore, such a condition can be incorporated into the  kinetic sector of the original action (\ref{Lag_Sym}) through a new dynamical variable (say undetermined Lagrange multiplier) $\lambda$, augmenting the initial Lagrangian density (\ref{Lag_Sym}) by the term $\lambda\dot{\Omega}$. However, it is better to include the constraint into the system in the form $-\dot{\lambda}\Omega$, instead of the $\lambda\dot{\Omega}$. The difference, being a full time-derivative, which does not modify the dynamics, that is
\begin{equation}
\lambda\dot{\Omega}=\frac{d}{dt}(\lambda\Omega)-\dot{\lambda}\Omega.
\end{equation}

In this way, discarding total time derivatives, we can write the first-iterated Lagrangian density as
\begin{equation}
{\mathcal{L}}^{(1)}=a_{I}(\xi)\dot{\xi}^{I}-\dot{\lambda}\Omega-\mathcal{H}^{(1)}(\xi),\label{LS0}
\end{equation}
where the first-iterated  Hamiltonian density is obtained by the relation
\begin{eqnarray}
\left.\mathcal{H}^{(1)}(\xi)=\mathcal{H}(\xi)\right\arrowvert_{\Omega= 0}.\label{Pot2}
\end{eqnarray}
It is worth remarking that the presence of $A_{0}$ has disappeared because the constraint (\ref{fj_constraint}) has been put into the kinetic part of the Lagrangian density (\ref{LS0}). In this way, by defining the new set of symplectic variables and the one-form respectively as
\begin{eqnarray}
{\xi}{^{(1)I}}&=& (A{_{a}},\Pi^{a}, \theta,P,\lambda),\label{variables}\\
{a}^{(1)}{_{I}} &=& (\Pi^{a},0, P,0,-\Omega);
\end{eqnarray}
we obtain the explicit expression of the first-iterated pre-symplectic matrix
\begin{equation}
\label{matrix2}
\mathcal{F}^{(1)}_{IJ}(\mathbf{x},\mathbf{y})=
\left(
  \begin{array}{cccccc}
  0   &  -\delta^{a}_{b}   &  0 &  0   &  - \mathcal{A}^{ac}\partial_{\mathbf{x}c} \\
 \delta^{a}_{b}  &   0      &   0   &    0   &  -\partial_{\mathbf{x}a} 	\\
 0   &   0    & 0  &  -1	 &  -\mathcal{B}^{c}\partial_{\mathbf{x}c} \\
 0  &    0  & 1  &  0 	& 0 	 \\
 \mathcal{A}^{bc}\partial_{\mathbf{y}c}  &  \partial_{\mathbf{y}b}  & \mathcal{B}^{c}\partial_{\mathbf{y}c}  &  0 	&  0
 \end{array}
\right)\delta^{3}(\mathbf{x}-\mathbf{y}).
\end{equation}
Here, we have defined $\mathcal{A}^{ac}\equiv\left(\kappa/8\pi^{2}\right)\epsilon^{0abc}P_{b}$ and $\mathcal{B}^{c}\equiv\left(\kappa/16\pi^{2}\right)\epsilon^{0abc}F_{ab}$. From the above singular matrix we can obtain the following non-trivial zero-mode:
\begin{equation}
v^{(1)I}= \left(-\partial^{\mathbf{y}}_{a}, \mathcal{A}^{ac}\partial^{\mathbf{y}}_{c} , 0, \mathcal{B}^{c}\partial^{\mathbf{y}}_{c} , -1 \right).\label{v1}
\end{equation}
However, if we multiply this zero-mode by the gradient of first-iterative Hamiltonian (\ref{Pot2}), we will find that there are no further constraints in the theory. This is an expected situation since this theory must have a local gauge symmetry. Following the symplectic framework, the zero-modes must act as the generators of the corresponding gauge symmetry `$\delta_G$' on the symplectic variables, that is, the components of the zero-mode should give the transformation properties related to the underlying symmetry through the whole phase space \cite{Wotzasek, Montani}. Thus, the local infinitesimal transformations, corresponding to the symplectic variables (\ref{variables}) and generated by $v^{(1)I}$ (\ref{v1}), can be written as
\begin{equation}
\delta_{G}\xi^{(1)}_{I}=v^{(1)}_{I}\eta,
\end{equation}
with $\eta$ an arbitrary infinitesimal parameter. From the above zero-mode (\ref{v1}), we can immediately obtain the following infinitesimal gauge transformations:
\begin{eqnarray}
\delta_{G}A_{a}&=&-\partial_{a}\eta,\\
\delta_{G}\Pi^{a}&=&\frac{\kappa}{8\pi^{2}}\epsilon^{0abc}P_{b}\partial_{c}\eta, \\
\delta_{G}\theta&=&0,\\
\delta_{G}P&=&\frac{\kappa}{16\pi^{2}}\epsilon^{0abc}F_{ab}\partial_{c}\eta,\\
\delta_{G}\lambda&=&-\eta.
\end{eqnarray}
In particular, we can identify $-\dot{\lambda}$ with $A_0$ such that all the physical quantities, i.e observables, are invariant under the following gauge transformations:
\begin{eqnarray}
A_{\mu}&\longrightarrow& A_{\mu}+\partial_{\mu}\eta,\label{gauge1}\\
\theta&\longrightarrow&\theta.\label{gauge2}
\end{eqnarray}
Note that the generator of such gauge transformation has been identified directly from the structure of the zero-mode (\ref{v1}), thereby making evident that the zero-mode of the pre-symplectic two-form matrix encode all the information of the gauge symmetry of our theory.

Now, to define the quantization brackets and also find the physical degrees of freedom, one can fix the gauge freedom associated to the singularity in the pre-symplectic matrix (\ref{matrix2}) by imposing additional conditions (gauge conditions). For electromagnetic fields, the condition usually employed to gauge fix is the Coulomb gauge. It satisfies the equation
\begin{equation}
\Phi=\partial_{a}A^{a}=0.\label{fixing2}
\end{equation}
Then, introducing this gauge condition into the Lagrangian density (\ref{LS0}) through the Lagrange multiplier $\alpha$, we arrive at the following second-iterated Lagrangian density:
\begin{equation}
{\mathcal{L}}^{(2)}(\xi, \dot{\xi})=a^{(1)}_{I}(\xi)\dot{\xi}^{(1)I}-\dot{\alpha}\Phi-\mathcal{H}^{(2)}(\xi),\label{LS}
\end{equation}
 where $\mathcal{H}^{(2)}$ is given by
\begin{eqnarray}
\mathcal{H}^{(2)}=\frac{1}{16\pi}F_{ab}F^{ab}-\frac{1}{2}P_{a}P^{a}-2\pi\left(\Pi^{a}-\frac{\kappa}{8\pi^{2}}\epsilon^{0abc}A_{b}P_{c}\right)^{2}+\frac{1}{2}\left(P+\frac{\kappa}{16\pi^{2}}\epsilon^{0abc}A_{a}F_{bc}\right)^{2}.
\end{eqnarray}
The Lagrangian density in (\ref{LS}) allows us read off the new set of symplectic variables, $\xi^{(2)}{_{I}}= (A{_{a}}, \Pi^{a},  \theta, P, \lambda, \alpha)$.  This allows us to identify the new canonical one-form, ${a}{^{(2)I}} =(\Pi^{a},0,P, 0,-\Omega, -\Phi)$. In this manner, using the pre-symplectic matrix definition in Eq. (\ref{symplectic_matrix}) again, we have the second-iterated matrix
\begin{equation}
\label{eq}
\mathcal{F}^{(2)}_{IJ}(\mathbf{x},\mathbf{y})=\left(
  \begin{array}{cccccc}
  0   &  -\delta^{a}_{b}   &  0 &  0   &  - \mathcal{A}^{ac}\partial_{\mathbf{x}c}&-\partial_{\mathbf{x}}^{a} \\
 \delta^{a}_{b}  &   0      &   0   &    0   &  -\partial_{\mathbf{x}}^{a} &0	\\
 0   &   0    & 0  &  -1	 &  -\mathcal{B}^{c}\partial_{\mathbf{x}c} &0\\
 0  &    0  & 1  &  0 	& 0 	&0 \\
 \mathcal{A}^{ac}\partial_{\mathbf{y}c}  &  \partial_{\mathbf{y}a}  & \mathcal{B}^{c}\partial_{\mathbf{y}c}  &  0 	&  0&0\\
 \partial_{\mathbf{y}a}	&	0	&	0	&	0	&0	&	0		\\
 \end{array}
\right)\delta^{3}(\mathbf{x}-\mathbf{y}).
\end{equation}
Using the standard identity for any matrix
\begin{eqnarray}
\label{det}
\left(
  \begin{array}{cc}
  \mathbf{A}	   & \mathbf{B}	  	  	\\
 \mathbf{C}  	 &	  \mathbf{D}     	\\
 \end{array}
\right)=\left(
  \begin{array}{cc}
  \mathbf{A}	   & {\bf 0}	  	  	\\
 \mathbf{C}  	 &	  {\bf 1}     	\\
 \end{array}
\right)\left(
  \begin{array}{cc}
  {\bf 1}	   & \mathbf{A}^{-1}\mathbf{B}	  	  	\\
 {\bf 0} 	 &	  \mathbf{D}-\mathbf{C}\mathbf{A}^{-1}\mathbf{B}     	\\
 \end{array}
\right),
\end{eqnarray}
where $\mathbf{A}$ and $\mathbf{D}$ are square matrices, but $\mathbf{B}$ and $\mathbf{C}$ need not be square, we can see that
\begin{equation}
\text{det}\, \left(
  \begin{array}{cc}
  \mathbf{A}	   & \mathbf{B}	  	  	\\
 \mathbf{C}  	 &	  \mathbf{D}     	\\
 \end{array}
\right)=\left(\text{det}\, \mathbf{A}\right)\left(\text{det}\,\left[\mathbf{D}-\mathbf{C}\mathbf{A}^{-1}\mathbf{B}\right]\right).\label{compute_det}
\end{equation}
Hence, making use of (\ref{compute_det}), and after a few algebra, it is possible to show that the determinant of the matrix $\mathcal{F}^{(2)}_{IJ}$ is
\begin{equation}
\left(\text{det}\,\mathcal{F}^{(2)}_{IJ}(\mathbf{x},\mathbf{y})\right)^{1/2}=\text{det}\left|\nabla^{2}\delta^{3}(\mathbf{x}-\mathbf{y})\right|\neq 0. \label{det}
\end{equation}
Thus, it is clear that $\mathcal{F}^{(2)}_{IJ}$ is not singular, and therefore the inverse of this matrix exists: it is dubbed as the symplectic two-form matrix. Now, it is interesting to note that according to Toms approach \cite{Toms}, the functional measure for the path integral associated with our model, in the physical Coulomb gauge, can be written out explicitly as
\begin{eqnarray}
d\mu&=&\left(\prod_{I} \left[D \xi^{(2)}_{I}\right]\right)\left(\text{det}\,\mathcal{F}^{(2)}_{IJ}(\mathbf{x},\mathbf{y})\right)^{1/2}\nonumber\\
&=&\left(\prod_{I} \left[D \xi^{(2)}_{I}\right]\right) \text{det}\left|\nabla^{2}\delta^{3}(\mathbf{x}-\mathbf{y})\right|.
\end{eqnarray}
A direct consequence of the last equation is that the quantum  transition amplitude can now be obtained by performing the integration over $D\lambda D\alpha$:
\begin{equation}
\mathcal{Z}=\int \left[DAD\Pi D\theta DP\right]\text{det}\left|\nabla^{2}\delta^{3}(\mathbf{x}-\mathbf{y})\right|\delta^{3}\left(\Omega\right)\delta^{3}\left(\Phi\right)\text{exp}\left\{\frac{i}{\hbar}\int \left(\dot{A}_{a}\Pi^{a}+\dot{\theta}P-\mathcal{H}^{(2)}\right)d^{4}x\right\}.
\end{equation}
According to the analysis in Refs. \cite{Liao,Huang}, one could use these results concretely to infer the functional generator of the Green function and/or connected Green function. This  allows us to have more information about the quantization of the system.

Also, after a straightforward calculation,  we obtain the symplectic two-form matrix,
\begin{equation}
\label{eq}
\left(\mathcal{F}^{(2)}_{IJ}(\mathbf{x},\mathbf{y})\right)^{-1}=
\left(
  \begin{array}{ccccccccc}
  0	   & \Delta_{a}^{b}(\mathbf{x})	  &	 0 	&	0   	& 0 	 &  	-\frac{\partial_{\mathbf{x}}^{ a}}{\nabla_{\mathbf{x}}^{2}}	  	\\
 -\Delta_{a}^{b}(\mathbf{y}) 	 &	  0      &	   0   &	  0 	 & -\frac{\partial_\mathbf{x}^{a}}{\nabla_{\mathbf{x}}^{2}}	&	 0		\\
 0 	  &	  0  	  & 0	  & 1		 & 	0 	&	0		\\
0  	&   0	&	 -1 	 &  0 	& 	0 	&	0		 \\
0  & \frac{\partial_\mathbf{y}^{a}}{\nabla_{\mathbf{y}}^{2}} 	 &	0	   &	0 	&	 0 	&		-\frac{1}{\nabla_{\mathbf{x}}^{2}}		 \\
\frac{\partial_{\mathbf{y}}^{a}}{\nabla_{\mathbf{y}}^{2}}	&	0	&	0	&	0	&\frac{1}{\nabla_{\mathbf{y}}^{2}}	&	0		\\
 \end{array}
\right)\delta^{3}(\mathbf{x}-\mathbf{y}),
\end{equation}
where  $\Delta_{a}^{b}(\mathbf{x})\equiv\left(\delta_{a}^{b}-\frac{\partial_{\mathbf{x}a}\partial_{\mathbf{x}}^{b}}{\nabla_{\mathbf{x}}^{2}}\right)$ is the transverse projective operator and $1/\nabla_{\mathbf{x}}^{2}$ is the inverse Laplacian operator. As a result, the form of the symplectic matrix (\ref{eq}) defines the brackets $\left\{\bullet\, ,\bullet\right\}_{\text{F-J}}$, dubbed as the  Faddev-Jackiw fundamental brackets, between any two elements of the symplectic variables set $\xi_{I}^{(2)}( \mathbf{x})$ over the phase space through
\begin{equation}
\left\{\xi_{I}^{(2)}(\mathbf{x}),\xi_{J}^{(2)}(\mathbf{y})\right\}_{\text{F-J}}=\left(\mathcal{F}_{IJ}^{(2)}(\mathbf{x},\mathbf{y})\right)^{-1}.
\end{equation}
In this way, we can read off the non-vanishing brackets between canonical variables
\begin{eqnarray}
\left\{A_{a}(\mathbf{x}),\Pi^{b}(\mathbf{y})\right\}_{\text{F-J}}&=&\Delta_{a}^{b}(\mathbf{x})\delta^{3}(\mathbf{x}-\mathbf{y}),\label{FJ1}\\
\left\{\theta(\mathbf{x}),P(\mathbf{y})\right\}_{\text{F-J}}&=&\delta^{3}(\mathbf{x}-\mathbf{y}).\label{FJ2}
\end{eqnarray}
 Then it is straightforward to see that the brackets (\ref{FJ1}) and (\ref{FJ2}) or equivalently the symplectic two-form (\ref{eq}) do not satisfy the common canonical commutation relations. However, they could allow us to study physical observables, that is, quantities that are involved in physical measurements such as work and energy \cite{, Allahverdyan,Henneaux,Dirac}. In this framework, the observables are objects forming zero brackets with all the theory's constraints.

In this respect, we now need a Poisson bracket for observables on $\Sigma$. Such a bracket should agree with the commutator in the classical limit. To be precise, for any two observables A, B defined on the phase space which owns itself a symplectic structure as $\left\{\xi_{I}^{(2)},\xi_{J}^{(2)}\right\}_{\text{F-J}}$, we can use the following relation
\begin{equation}
\{A(\xi),B(\xi)\}=\sum_{I,J}\int d^{3}\mathbf{r}\frac{\delta A(\xi)}{\delta\xi_{I}(\mathbf{r})}\left(\mathcal{F}_{IJ}^{(2)}\right)^{-1}\frac{\delta B(\xi)}{\delta\xi_{J}(\mathbf{r})}.
\end{equation}
This can be taken as the definition of the Poisson bracket.

The canonical quantization can be fully made at tree level by the replacement of classical observables and Poisson brackets by the quantum operators commutators, respectively, according to
\begin{equation}
\left\{A(\mathbf{x}),B(\mathbf{y})\right\}\longrightarrow\frac{1}{i\hbar}\left[ \hat{A}(\mathbf{x}),\hat{B}(\mathbf{y})\right],\,\hat{\mathcal{O}}|\psi\rangle=0,
\end{equation}
where $\hat{\mathcal{O}}$ is any operator associated with an observable (or constraint) and $|\psi\rangle$ is any quantum state.

 Finally, the number of propagating degrees of freedom may be calculated in the phase space from the relation
\begin{equation}
N=\frac{1}{2}\left(N_{1}-N_{2}-N_{3}\right),
\end{equation}
where $N_{1}$ is the number of field components in $\xi_{I}=(A{_{0}},A{_{a}},\Pi^{a}, \theta,P)$, $N_{2}$ is the number of  fields eliminated $(A_{0})$, and $N_{3}$ is the number constraints including gauge fixing conditions. Hence,  it is concluded that contrary to standard electrodynamics, axion electrodynamics has $\frac{1}{2}\left(9-1-2\right)$= 3 physical degrees of freedom on $\Sigma$; two degrees of freedom  for the photon and another for the axion.

\section{Summary and conclusion}
\label{final}
In this work, we have studied the conservation laws and dynamical structure of $(3 + 1)-$dimensional modified axion electrodynamics theory. We have started by reviewing some known results that exist in the literature concerning field equations. Thereafter, using the covariant phase space method, we have built the Noether charges associated with the gauge transformations of the theory. Such charges turned out to be the sum of the Noether electric charge plus a \textit{magnetic} charge induced by the dynamical axion field. Furthermore, we have already inferred the improved energy-momentum tensor by adding a superpotential term to the canonical energy-momentum tensor, which, however,  remains antisymmetric making evident the absence of Lorentz invariance in the theory. After constructing the improved energy-momentum tensor, we have determined the expressions for the energy density, momentum density, and energy flux density, respectively. Interestingly, these expressions are not gauge invariant due to  $\mathbf{A}$.

Furthermore, in the Faddev-Jackiw symplectic framework, we have derived the transformation laws for all the set of dynamical variables corresponding to gauge symmetries of the system, which have been used to compute the conserved charges. Additionally, we have used an appropriate gauge-fixing procedure, the Coulomb gauge, to compute the quantization brackets, which  are useful for studying physical observables; e.g., the interaction energy  between static point-like sources and the thermodynamics work done on the electromagnetic field coupled to axion fields  via moving charges \cite{Allahverdyan}. In this setup, we have also obtained the functional measure on the path integral associated with our model. A direct consequence of this is that the quantum transition amplitude expression can now be constructed by following the procedure developed in Ref. \cite{Toms}. Having these results  at hand,   one might be faced with the task of calculating the generating functionals of the Green function and connected Green function,  which could get us more information on the quantization of the theory. This idea is in progress and will be the subject of forthcoming work. We conclude our study by confirming that axion electrodynamics has  three physical degrees of freedom per space point: two degrees of freedom  for the photon and another for the axion. One of the interesting features of this analysis is the fact that both the gauge symmetry and the constraint structure of this model have been directly extracted by studying only the properties of the pre-symplectic matrix and its corresponding zero-modes.

\section{Acknowledgments}
This work has been supported by the National Council of Science and Technology,  and Sistema Nacional de Investigadores (M\'exico). We would like to thank G. Tavares-Velasco and Rodrigo Olea for useful discussions and suggestions.


\begin{thebibliography}{99}

\bibitem{Peccei} R. D. Peccei, H. R. Quinn, \textit{CP Conservation in the Presence of Pseudoparticles}, Phys. Rev. Lett. 38  1440 (1977).
\bibitem{Peccei2}R. D. Peccei, \textit{The Strong CP problem and axions}, Lect. Notes Phys. 741 (2008) 3-17.
\bibitem{Wilczek0}F. Wilczek,\textit{Problem of Strong  $P$  and  $T$  Invariance in the Presence of Instantons}, Phys.Rev.Lett. 40 (1978) 279-282.
\bibitem{Wilczek} F. Wilczek,\textit{Two applications of axion electrodynamics}, Phys. Rev. Lett. 58, 1799 (1987).
\bibitem{Weinberg} S. Weinberg, \textit{A New Light Boson?}, Phys. Rev. Lett. 40 (1978) 223.
\bibitem{Kim}J. E. Kim, \textit{Weak Interaction Singlet and Strong CP Invariance}, Phys.Rev.Lett. 43 (1979) 103.
\bibitem{Shifman}M. A. Shifman, A. I. Vainshtein and V. I. Zakharov, \textit{Can Confinement Ensure Natural CP Invariance of Strong Interactions?}, Nucl.Phys.B 166 (1980) 493-506.
\bibitem{Dine}M. Dine, W. Fischler and M. Srednicki, \textit{A Simple Solution to the Strong CP Problem with a Harmless Axion}, Phys.Lett.B 104 (1981) 199-202.
\bibitem{Zhitnitsky}A. R. Zhitnitsky, \textit{On Possible Suppression of the Axion Hadron Interactions.}, Sov.J.Nucl.Phys. 31 (1980) 260, [Yad. Fiz.31,497(1980)].
\bibitem{Jihn} Jihn E. Kim and Gianpaolo Carosi, \textit{Axions and the Strong CP Problem}, Rev. Mod. Phys. 82 (2010) 557-602.
\bibitem{axion_dark}Laura Covi, Hang-Bae Kim, Jihn E. Kim and Leszek Roszkowski, \textit{Axinos as dark matter}, JHEP 0105 (2001) 033.
\bibitem{Bradley}R. Bradley et al., \textit{Microwave cavity searches for dark-matter axions}, Rev. Mod. Phys. 75, 777 (2003).
\bibitem{Vergados}J. D. Vergados and Y. K. Semertzidis, \textit{Axionic dark matter signatures in various halo model}, Nucl.Phys. B 915 (2017) 10-18.
\bibitem{Camila} Camila S. Machado, Wolfram Ratzinger, Pedro Schwaller and Ben A. Stefanek, \textit{Audible axions}, JHEP 01 (2019) 053.
\bibitem{Lombardo}M. P. Lombardo and A. Trunin, \textit{Topology and axions in QCD}, Int. J. Mod. Phys. A 35, no.20, 2030010 (2020).
\bibitem {Preskill}J. Preskill, M.B. Wise and F. Wilczek, \textit{Cosmology of the Invisible Axion}, Phys. Lett. B 120 (1983) 127.
\bibitem{Nelson}A. E. Nelson and J. Scholtz, \textit{Dark Light, Dark Matter and the Misalignment Mechanism}, Phys. Rev. D 84 (2011) 103501.
\bibitem{Raymond} Raymond T. Co, Lawrence J. Hall and Keisuke Harigaya, \textit{Axion Kinetic Misalignment Mechanism}, Phys. Rev. Lett., 124, 251802 (2020).
\bibitem{Arvanitaki}A. Arvanitaki, S. Dimopoulos, S. Dubovsky, N. Kaloper and J. March-Russell, \textit{String Axiverse}, Phys. Rev. D 81, 123530 (2010).
\bibitem{Raffelt} Georg G. Raffelt, \textit{Stars as laboratories for fundamental Physics}, University of Chicago Press, 1996; Lectures Notes Physics 741, 51 (2008).
\bibitem{Qi0}X. -L. Qi, T. L. Hughes, S. -C. Zhang, \textit{Topological field theory of time-reversal invariant insulators}, Phys. Rev. B 78 (2008) 195424.
\bibitem{Rundong}Rundong Li, Jing Wang, Xiao-Liang Qi and Shou-Cheng Zhang, \textit{Dynamical axion field in topological magnetic insulators}, Nature Phys 6, 284-288 (2010).
\bibitem{Akihiko}Akihiko Sekine and Kentaro Nomura, \textit{Axion electrodynamics in topological materials},	J. Appl. Phys. 129, 141101 (2021).
\bibitem{Nenno}Nenno, D.M., Garcia, C.A.C., Gooth, J. et al., \textit{Axion physics in condensed-matter systems}, Nat Rev Phys 2, 682–696 (2020).
\bibitem{Katsuhisa}Katsuhisa Taguchi et al., \textit{Electromagnetic effects induced by time-dependent axion field}, Phys. 
Rev. B 97, 214409 (2018).
\bibitem{DMEF}Jinlong Zhang, Dinghui Wang, Minji Shi, Tongshuai Zhu, Haijun Zhang, and Jing Wang, \textit{ Large dynamical axion field in topological antiferromagnetic insulator Mn$_2$Bi$_2$Te$_5$}, Chin. Phys. Lett. 37, 077304 (2020), [arXiv:1906.07891].
\bibitem{Xia}Xia, Y., Qian, D., Hsieh, D. et al., \textit{Observation of a large-gap topological-insulator class with a single Dirac cone on the surface}, Nature Phys 5, 398-402 (2009).
\bibitem{Zhang}Zhang, H., Liu, C., Qi, X. et al, Topological insulators in $Bi_2Se_3$, $Bi_2Te_3$ and $Sb_2Te_3$ with a single Dirac cone on the surface, Nature Phys 5, 438–442 (2009).
\bibitem{Tokura} Tokura, Y., Yasuda, K. and Tsukazaki, A, \textit{Magnetic topological insulators}, Nat. Rev. Phys. 1, 126-143 (2019).
\bibitem{Wilczek2} Wilczek, F. \textit{A theoretical physicist examines exotic particles lurking in new materials}, Journal club. Nature 458, 129 (2009).
\bibitem{Sikivie} P. Sikivie, \textit{Experimental Tests of the “Invisible” Axion},  Phys.Rev.Lett. 50 (1983) 1415, Phys.Rev.Lett. 51 (1983) 1395 (erratum).
\bibitem{Sikivie1}P. Sikivie, \textit{Detection rates for “invisible”-axion searches}, Phys. Rev. D 32 (1985) 2988–2991.
\bibitem{Primakoff}H. Primakoff, Phys. Rev. 81, 899 (1951).
\bibitem{Stodolsky}G. Raffelt and L. Stodolsky, \textit{Mixing of the Photon with Low Mass Particles}, Phys. Rev. D 37 (1988) 1237.
\bibitem{Igor}Igor G. Irastorza and Javier Redondo, \textit{New experimental approaches in the search for axion-like particles}, Prog.Part.Nucl.Phys. 102 (2018) 89-159.
\bibitem{ADMX} ADMX collaboration, \textit{A Search for Invisible Axion Dark Matter with the Axion Dark Matter Experiment}, Phys. Rev. Lett. 120 (2018) 151301.
\bibitem{HAYSTAC}  S. Al Kenany, et al., Nucl. Instrum. Methods A 854 (2017) 11–24.
\bibitem{CULTASK} C. Woohyun, CULTASK, \textit{the Coldest Axion Experiment at CAPP/IBS in Korea}, PoS ICHEP 2016 (2016) 197.
\bibitem{MADMAX} MADMAX Working Group collaboration, \textit{Dielectric haloscopes: a new way to detect axion dark matter}, Phys. Rev. Lett. 118 (2017) 091801.
\bibitem{ABRACADABRA} Y. Kahn, B.R. Safdi, J. Thaler, Phys. Rev. Lett. 117 (14) (2016) 141801.
\bibitem{Orpheus}G. Rybka, A. Wagner, A. Brill, K. Ramos, R. Percival and K. Patel, \textit{Search for dark matter axions with the Orpheus experiment}, Phys. Rev. D 91 (2015) 011701.
\bibitem{ORGAN} B.T. McAllister, G. Flower, E.N. Ivanov, M. Goryachev, J. Bourhill and M.E. Tobar, \textit{The ORGAN Experiment: An axion haloscope above 15 GHz}, Phys. Dark Univ. 18 (2017) 67.
\bibitem{ORGAN2} B.T. McAllister, G. Flower, L.E. Tobar and M.E. Tobar, \textit{Tunable Supermode Dielectric Resonators for Axion Dark-Matter Haloscopes}, Phys. Rev. Applied 9 (2018) 014028.
\bibitem{RADES} Alejandro Alvarez Melcon et al. \textit{Axion Searches with Microwave Filters: the RADES project}. JCAP, 1805 (05):040, 2018. 
\bibitem{David}David J. E. Marsh, Kin Chung Fong, Erik W. Lentz, Libor Smejkal, and Mazhar N. Ali, \textit{Proposal to Detect Dark Matter using Axionic Topological Antiferromagnets}, Phys.Rev.Lett. 123 (2019) no.12, 121601.
\bibitem{Engel}Sch\"utte-Engel, Jan and Marsh, David J. E. and Millar, Alexander J. and Sekine, Akihiko and Chadha-Day, Francesca and Hoof, Sebastian and Ali, Mazhar and Fong, Kin-Chung and Hardy, Edward and \v{S}mejkal, Libor, \textit{ Axion Quasiparticles for Axion Dark Matter Detection}, arXiv: 2102.05366.
\bibitem{Josh} Josh Kirklin, \textit{Unambiguous Phase Spaces for Subregions}, JHEP 03 (2019) 116.
\bibitem{Allahverdyan}A.E. Allahverdyan, D. Karakhanyan, \textit{Defining the Work Done on an Electromagnetic Field}, Phys.Rev.Lett. 121 (2018) 24, 240602.
\bibitem{Henneaux}M. Henneaux and C. Teitelboim, \textit{Quantization of Gauge Systems}, Princeton, New Jersey: Princeton
University Press 1991.
\bibitem{Noether} E. Noether, \textit{Invariant Variation Problems}, Gott. Nachr. 1918 (1918) 235 [Transp. Theory Statist. Phys. 1 (1971) 186] doi:10.1080/00411457108231446 [physics/0503066].
\bibitem{Witten}C. Crnković and E. Witten, \textit{Covariant Description of Canonical Formalism in Geometrical Theories}, in Three Hundred Years of Gravitation, edited by W. Israel and S. W. Hawking (Cambridge University Press, 1987).
\bibitem{Noether-Wald}J. Lee and Robert M. Wald, \textit{Local symmetries and constraints}, J. Math. Phys. 31, 725-743 (1990).
\bibitem{Wald} Vivek Iyer and Robert M. Wald, \textit{Some properties of Noether charge and  proposal for dynamical black hole entropy},” Phys.Rev., D50, 846–864 (1994).
\bibitem{F-J}L.D. Faddeev, R. Jackiw, \textit{Hamiltonian Reduction of Unconstrained and Constrained Systems}, Phys. Rev. Lett. 60, 1692 (1988).
\bibitem{Wotzasek} J. Barcelos-Neto, C. Wotzasek, \textit{Faddeev-Jackiw quantization and constraints}, Int. J. Mod. Phys. A 7, (1992) 4981.
\bibitem{Montani}H. Montani and R. Montemayor, \textit{Lagrangian approach to a symplectic formalism for singular systems}, Phys. Rev. D 58 (1998) 125018.
\bibitem{Toms}  D. J. Toms, \textit{Faddeev-Jackiw quantization and the path integral}, Phys. Rev. D 92 105026 (2015).
\bibitem{Dirac}P. A. M. Dirac, \textit{Lecture on quantum mechanics}, Yeshiva University, U.S.A. (1964).
\bibitem{Belinfante2}F. J. Belinfante, \textit{On the current and the density of the electric charge, the energy, the linear momentum and the angular momentum of arbitrary fields}, Physica {\bf 7} (1940) 449.
\bibitem{Rosenfeld}L. Rosenfeld, \textit{Sur le tenseur d'impulsion-\'energy}, Mem. Acad. Roy. Belg. Sci. {\bf 18} (1940) 1.
\bibitem{Daniel}Daniel N. Blaschke, Franois Gieres, M\'eril Reboud, and Manfred Schweda, \textit{The energy-momentum tensor(s) in classical gauge theories}, Nucl.Phys. {\bf B} 912, 192-223 (2016).
\bibitem{PDG}P.A. Zyla et al. [Particle Data Group], PTEP 2020, no.8, 083C01 (2020) doi:10.1093/ptep/ptaa104.
\bibitem{Carrol}S. M. Carroll, G. B. Field, and R. Jackiw, \textit{Limit on a Lorentz- and parity-violating modification of electrodynamics}, Phys. Rev. D41, 1231 (1990).
\bibitem{Dunne}Gokce Basar and Gerald V. Dunne, \textit{The Chiral Magnetic Effect and Axial Anomalies}, Lect.Notes Phys. 871 (2013) 261-294.
\bibitem{Dmitri}Dmitri E. Kharzeev,\textit{Topologically induced local P and CP violation in QCD$\times$QED},  Annals Phys. 325 (2010) 205-218.
\bibitem{Tobar}Michael E. Tobar, Ben T. McAllister and Maxim Goryachev, \textit{Modified axion electrodynamics as impressed electromagnetic sources through oscillating background polarization and magnetization},  Phys.Dark Univ. 26 (2019) 100339.
\bibitem{Naoto}Naoto Nagaosa, Jairo Sinova, Shigeki Onoda, A. H. MacDonald, and N. P. Ong, \textit{Anomalous Hall effect}, Rev. Mod. Phys. 82, 1539 (2010).
\bibitem{OmarJ}O. J. Franca, L. F. Urrutia, Omar Rodr\'iguez-Tzompantzi, \textit{Reversed electromagnetic Vavilov-\v{C}erenkov radiation in naturally existing magnetoelectric media}, Phys. Rev. D {\bf 99} 116020 (2019).
\bibitem{Alberto} A. Mart\'in-Ruiz, M. Cambiaso and L. F. Urrutia, \textit{Green's function approach to Chern-Simons extended electrodynamics: An effective theory describing topological insulators}, Phys. Rev. {\bf D} 92, 125015 (2015).
\bibitem{witten} Witten, E. \textit{Dyons of charge E-$\theta$/2p}. Phys. Lett. B86, 283–287 (1979).
\bibitem{Liao}Leng Liao and Yong-Chang Huang, \textit{Path integral quantization corresponding to Faddeev-Jackiw canonical quantization},  Phys.Rev.D 75 (2007) 025025.
\bibitem{Huang}Huang, Yong-Chang and Liao, Leng and Lee, Xie-Guo, \textit{Faddeev-Jackiw canonical path integral quantization for a general scenario, its proper vertices and generating functionals}, Eur.Phys.J.C 60 (2009) 481-487.
\end{thebibliography}
\end{document}